\documentclass[aps,prd, twoside,superscriptaddress,showpacs, preprintnumbers, showpacs, nofootinbib, 
twocolumn]{revtex4-1}

\pdfoutput=1
\usepackage{amssymb}
\usepackage{amsthm}
\usepackage{graphicx,amsmath}
\usepackage{color}
\newcommand{\beq}{\begin{eqnarray}}
\newcommand{\eeq}{\end{eqnarray}}

\newcommand{\bmp}{\noindent\begin{minipage}{16cm}}
\newcommand{\emp}{\end{minipage}\vskip 7mm} 

\usepackage{dcolumn}
\usepackage{bm}
\usepackage{bbm}
\usepackage{subfigure}
\usepackage{slashed} 
\usepackage{mathrsfs}

\usepackage{epsfig}

\usepackage{hyperref}

\newcommand{\bea}{\begin{eqnarray}}
\newcommand{\eea}{\end{eqnarray}}

\newcommand{\ba}{\begin{eqnarray}}
\newcommand{\ea}{\end{eqnarray}}

\newcommand{\be}{\begin{eqnarray}}
\newcommand{\ee}{\end{eqnarray}}

\begin{document}


\title{Annihilating Asymmetric Dark Matter}
%

\author{Nicole F. Bell}
\email{n.bell@unimelb.edu.au}
\affiliation{ARC Centre of Excellence for Particle Physics at the Terascale, \\
School of Physics, The University of Melbourne, Victoria 3010, Australia}
\author{Shunsaku Horiuchi}
\email{horiuchi@vt.edu} 
\affiliation{Center for Neutrino Physics, Department of Physics, Virginia Tech, 
Blacksburg, VA 24061, USA}
\affiliation{Center for Cosmology, Department of Physics and Astronomy,
University of California, Irvine, CA 92697-4575, USA}
\author{Ian M. Shoemaker}
\email{shoemaker@cp3-origins.net} 
\affiliation{CP$^{3}$-Origins \& Danish Institute for Advanced Study  DIAS, University of Southern Denmark, Campusvej 55, DK-5230 Odense M, Denmark}


\begin{abstract}
The relic abundance of particle and antiparticle dark matter (DM) need not be vastly different in thermal asymmetric dark matter (ADM) models. By considering the effect of a primordial asymmetry on the thermal Boltzmann evolution of coupled DM and anti-DM, we derive the requisite annihilation cross section. This is used in conjunction with CMB and {\it Fermi}-LAT gamma-ray data to impose a limit on the number density of anti-DM particles surviving thermal freeze-out. When the extended gamma-ray emission from the Galactic Center is reanalyzed in a thermal ADM framework, we find that annihilation into $\tau$ leptons prefer anti-DM number densities 1-4$\%$ that of DM while the $b$-quark channel prefers 50-100$\%$. 
\\

{\footnotesize  \it Preprint: CP3-Origins-2014-029 DNRF90, DIAS-2014-29}
 \end{abstract}

\maketitle

\section{Introduction}


 The dominant contribution to the matter density of the Universe is non-luminous and has so far only been conclusively detected via its gravitational interactions. The weakly-interacting massive particle (WIMP) paradigm~\cite{Bertone:2004pz} postulates that in the early Universe this dark matter (DM) was kept in equilibrium with the thermal bath via non-gravitational interactions, and that weak-scale annihilation cross sections yield a thermal relic with the correct observed abundance of DM. The WIMP paradigm thus gives hope for a non-gravitational detection of this DM in three avenues: (1) producing it at a collider, $\bar{f}f \rightarrow \bar{X}X$, where $X$ is DM and $f$ is a Standard Model (SM) particle; (2) direct detection of the recoil imparted to a SM particle from scattering, $X f \rightarrow Xf$;  and (3) the indirect detection of DM annihilating to SM final states, $\bar{X}X \rightarrow \bar{f}f$. This final possibility is the only one requiring the presence of both particle $X$ and antiparticle $\bar{X}$ (having equal densities in the WIMP picture).

However, the fact that the abundances of the dark and the baryonic matter densities are within a factor of a few of each other could be an indication of a common origin. This curious fact has stimulated a substantial body of theoretical work in which DM carries a particle/antiparticle asymmetry (see~\cite{Petraki:2013wwa,Zurek:2013wia} for reviews). A feature common to many such models is the need to ensure a sufficiently large annihilation cross section such that the symmetric component does not yield too large a mass density. However, in the presence of an asymmetry, the thermal Boltzmann evolution of DM reveals that the symmetric component of DM need not be small~\cite{Griest:1986yu,Belyaev:2010kp,Graesser:2011wi} and may thus render ADM amenable to indirect detection searches~\cite{Graesser:2011wi}.

Annihilation signals of DM are unique in their sensitivity to the present density of anti-DM.  Given that ADM annihilation rates are suppressed by the small anti-DM fraction, $n_{\bar{X}}$, this requires a concomitantly larger annihilation cross section relative to the symmetric WIMP case.  This implies a fundamental degeneracy in indirect searches between the annihilation cross section, $\langle \sigma v \rangle$, and the number density of anti-DM, $n_{\bar{X}}$.  This degeneracy can be broken however by requiring that $\langle \sigma v \rangle$ be sufficient to account for the correct thermal relic abundance of $X$ and $\bar{X}$.  Thus with the input of an annihilation signal from DM one can infer both the annihilation cross section and the DM asymmetry, with direct consequences for DM model-building and phenomenology.

Such a signal may already have been found: the extended gamma-ray emission from the Galactic Center is consistent with annihilating DM in the $\sim 10-40$ GeV mass range~\cite{Goodenough:2009gk,Hooper:2010mq,Hooper:2011ti,Hooper:2011ti,Abazajian:2012pn,Gordon:2013vta,Macias:2013vya,Abazajian:2014fta,Daylan:2014rsa}. This has led to a large body of theoretical work to account for the observed signal while remaining consistent with collider and direct searches~\cite{Hardy:2014dea,Finkbeiner:2014sja,Lacroix:2014eea,Alves:2014yha,Berlin:2014tja,Agrawal:2014una,Izaguirre:2014vva,Cerdeno:2014cda,Ipek:2014gua,Kong:2014haa,Ko:2014gha,Boehm:2014bia,Abdullah:2014lla,Ghosh:2014pwa,Martin:2014sxa,Berlin:2014pya,Basak:2014sza,Cline:2014dwa,Detmold:2014qqa,Wang:2014elb,Arina:2014yna,Cheung:2014lqa,Huang:2014cla,Ko:2014loa,Okada:2014usa}.  We leave the analysis of a complete model for the Galactic Center in the light of ADM for future work.
 
In this paper we study the model-independent impact of indirect searches on thermal ADM. In Section II we describe the limits obtained from the gamma-ray data of the {\it Fermi}-Large Area Telescope (LAT) satellite and the cosmic microwave background (CMB), as well as the possible signal of annihilating DM from the Galactic Center. In Section III we derive the key requirements for thermal ADM and show how to map indirect constraints and signals into the parameters relevant for ADM.  We conclude in Sec. IV.

\section{Annihilation Constraints}
DM particles pair-annihilate into SM particles with model-dependent final states and branching ratios. Since the interaction involves a pair of particles, the annihilation rate scales as the DM density squared, motivating searches in regions of high DM density. Strong limits have been placed by a variety of searches using various DM sources and messenger particles. In some cases, excess signals remain after subtracting known backgrounds, leading to DM interpretations. In this section we summarize the current status of limits and detections, and discuss implications in the next section. 

The flux of particle species, $i$, from DM self-annihilation generally follows the form \cite{Bertone:2004pz},
\begin{equation}
\phi_i(E) = \frac{1}{4 \pi m_{X}^2} \langle \sigma_{\rm ann} v_{\rm rel} \rangle \frac{dN_i}{dE} \int_{l.o.s.} ds \rho^2_{DM}(r),
\end{equation}
where $m_{X}$ is the DM mass, $\langle \sigma_{\rm ann} v_{\rm rel} \rangle$ is the averaged annihilation cross section multiplied by the relative velocity, $dN_i/dE$ is the spectrum of particles $i$ resulting from each annihilation, and the integral takes the line of sight integral of the DM density squared. In the absence of a detection, observational data set upper limits on the size of $\phi_i(E)$ that can come from DM annihilation and still be compatible with data, which can then be interpreted as a limit on $\langle \sigma_{\rm ann} v_{\rm rel} \rangle$ as a function of the DM mass. Complicating this process is the presence of non-DM foregrounds and backgrounds that must be modeled and subtracted. Additional uncertainties arise since $dN_i/dE$ can be affected by astrophysics, in particular for charged particles that can rapidly lose energy and/or propagate diffusively depending on the environmental parameters. Multiple sources, methods, and analyses help break some degeneracies \cite{Ng:2013xha}. 

\begin{figure}[t]
\begin{center}
\includegraphics[width=0.45\textwidth,height=0.45\textwidth]{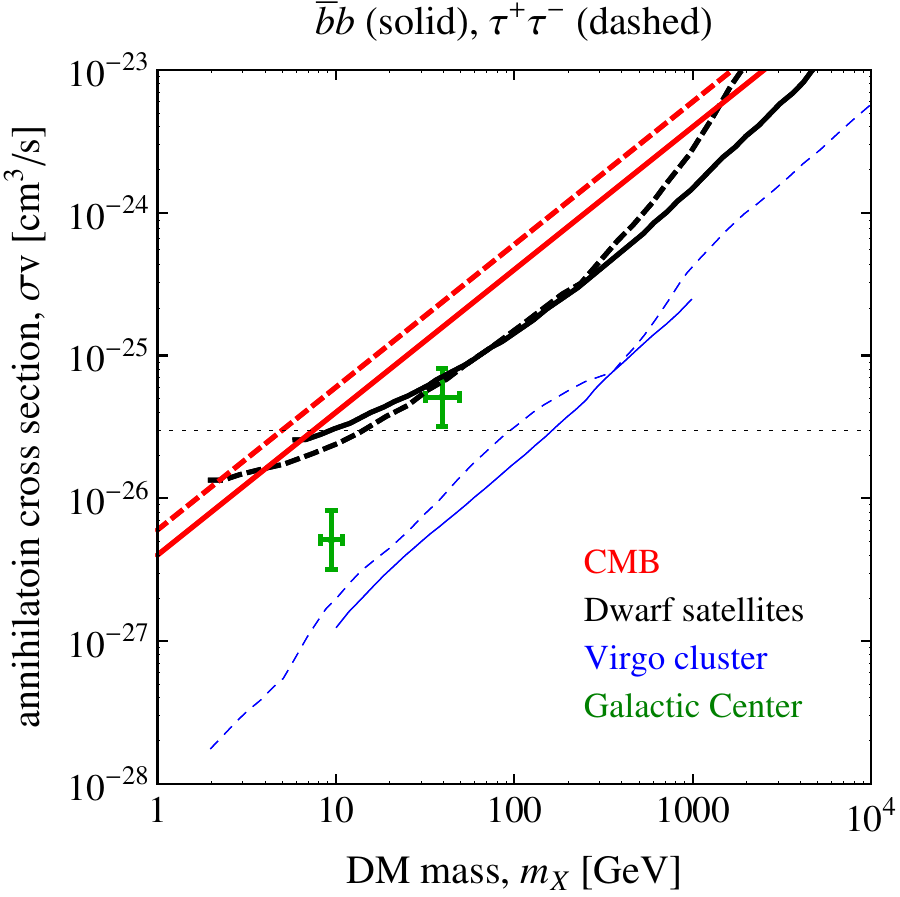}
\caption{95\% CL upper limits on the DM annihilation cross section from the CMB (bold red), satellite galaxies of the Milky Way (bold black), and the Virgo cluster (thin blue). The solid lines are for a $100$\% $b\bar{b}$ annihilation channel, while the dashed lines are for $\tau^+\tau^-$. These limits are obtained under the assumption of self-annihilating DM. The green data points denote the DM parameters that can account for the extended gamma-ray emission from the Galactic Center. The bold lines denote robust limits, while the thin lines denote limits which are subject to larger systematic uncertainties.}
\label{fig:limits}
\end{center}
\end{figure}

We focus on two constraints, one arising from searches using gamma-ray data from the Fermi-LAT satellite, and the other using the indirect effects of DM annihilation on the CMB. Figure \ref{fig:limits} summarizes constraints arising from searches in satellite dwarf galaxies of the Milky Way \cite{Ackermann:2013yva}, nearby galaxy clusters \cite{Han:2012uw}, and analysis of combined WMAP, Planck, ACT, SPT, BAO, HST, and SN data \cite{Madhavacheril:2013cna}. Only the annihilation channel $100$\% $b\bar{b}$ (solid lines) or $\tau^+\tau^-$ (dashed lines) are show for clarify. A variety of additional constraints, not shown, are within the range of limits shown in the Figure. These are from, e.g., analysis of the Milky Way halo \cite{Ackermann:2012rg}, the extragalactic diffuse gamma-ray emission \cite{Abazajian:2010sq,Abdo:2010dk}, analysis of gamma-ray anisotropies \cite{Ando:2013ff}, and cross-correlations with distribution of DM structures \cite{Shirasaki:2014noa}. 

The dwarf limits of Ref.~\cite{Ackermann:2013yva} combines Fermi-LAT observations in the directions of a total of $15$ satellites galaxies of the Milky Way, and are an update of previous estimates \cite{Abdo:2010ex,GeringerSameth:2011iw,Ackermann:2011wa} in both data and analysis. The closest dwarfs are treated as extended sources, kinematic data are used to estimate the DM contents of the galaxies, and a joint likelihood is performed marginalizing over other parameters. Due to the negligible astrophysical backgrounds in these galaxies, the DM limits are among the most robust. The cluster limits of Ref.~\cite{Han:2012uw} provide stronger limits, but additional uncertainties in the precise DM content, density profiles, and contaminations from cosmic-ray induced gamma-ray emission exist. We elect to adopt the limits from the Virgo cluster, which are among the strongest, and show the most conservative limit that does not subtract a cosmic-ray contribution. Nevertheless, the limits are susceptible to significant systematic uncertainty. For example, \cite{Han:2012uw} uses a boost factor of $\sim 10^3$; more recent estimates suggest values closer to $\sim 35$ \cite{Sanchez-Conde:2013yxa}, which would weaken the cluster limits by a factor of $\sim 30$. 

The CMB limits of Ref.~\cite{Madhavacheril:2013cna} constraints the energy deposition at very high redshifts ($1400 > z > 100$) by using the temperature and polarization anisotropies of the CMB.  The current (WMAP9+Planck+ACT+SPT+BAO+HST+SN) 95$\%$ CL limit on self-annihilating DM is $f \langle \sigma v \rangle_{SDM} /m_{X} < 1.2 \times 10^{-27}~{\rm cm}^{3}{\rm s}^{-1}{\rm GeV}^{-1}$~\cite{Madhavacheril:2013cna}, where $f$ is a function that describes the fraction of energy that is absorbed by the CMB plasma. The limits are particularly powerful for light-mass DM candidates \cite{Galli:2011rz,Hutsi:2011vx,Finkbeiner:2011dx,Lin:2011gj}, and uncertainties of the energy deposition efficiently is now small \cite{Madhavacheril:2013cna,Weniger:2013hja}; we adopt $f=0.3$ for the $b\bar{b}$ channel and $0.2$ for the $\tau^+\tau^-$ channel.

Towards the Galactic Center (GC) of our Milky Way, the situation is driven more complex by intense astrophysical backgrounds. Emissions from cosmic-ray interactions with target gas dominates the observed gamma rays and is complex to model. In addition, unresolved point sources such as milli-second pulsars can mimic an extended gamma-ray source like DM annihilation \cite{Baltz:2006sv,Abazajian:2010zy} (but, see, e.g., Refs.~\cite{Hooper:2013nhl,Cholis:2014lta}). Despite the complexity, several groups have found strong evidence for an extended emission source in Fermi-LAT data that is consistent with the spatial distribution expected in DM halo formation simulations, a spectrum that is consistent with the annihilation of $\sim 10$--$30$ GeV DM to quarks or leptons, and a flux that is consistent with the annihilation rate of thermally produced WIMP DM \cite{Goodenough:2009gk,Hooper:2010mq,Hooper:2011ti,Hooper:2011ti,Abazajian:2012pn,Gordon:2013vta,Macias:2013vya,Abazajian:2014fta,Daylan:2014rsa}. When the extended gamma-ray emission is interpreted as being due to DM annihilation, it favors $m_{X} = 39.4$ GeV for annihilation in $b$ quarks with an annihilation cross section $(5.1\pm 2.4)\times 10^{-26}~{\rm cm}^{3}{\rm s}^{-1}$, and a 9.4 GeV DM mass for annihilation into $\tau$ leptons with cross section $(0.51\pm .24)\times 10^{-26}~{\rm cm}^{3}{\rm s}^{-1}$~\cite{Abazajian:2014fta}. As there is no strong evidence for a preference for one channel over the other, we consider both separately. 

\section{Limits on Asymmetric DM}

\begin{figure}[t]
\begin{center}
\includegraphics[width=.45\textwidth, height=0.45\textwidth]{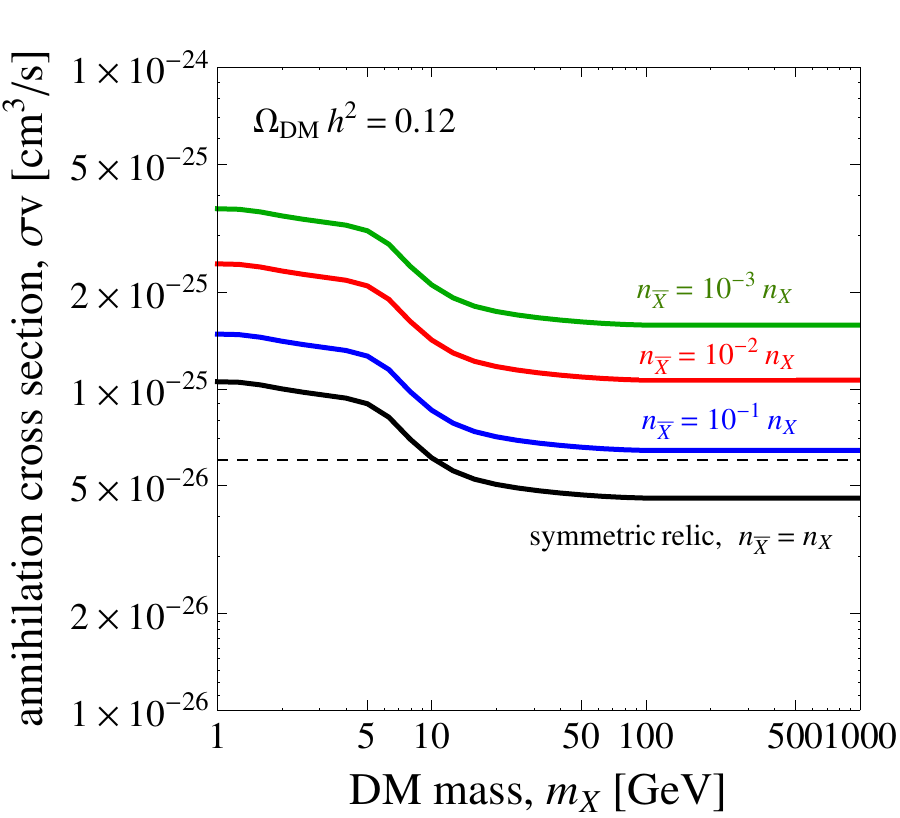}
\caption{Requisite total annihilation cross section for the symmetric $n_{\overline{X}} = n_{X}$ case along with the asymmetric cases $n_{\overline{X}} = 10^{-1} n_{X}$, $10^{-2} n_{X}$, and $10^{-3} n_{X}$.  For reference the canonical WIMP cross section for Dirac fermion or complex scalar DM is shown as the dashed curve, $\langle \sigma v \rangle = 6 \times 10^{-26}~{\rm cm}^{3}{\rm s}^{-1}$.}
\label{fig:rinfrelic}
\end{center}
\end{figure}
In the conventional symmetric thermal relic scenario, the abundance of DM is determined by the annihilation cross section $\langle \sigma_{ann} v_{rel} \rangle$ at the time of thermal freeze-out~\footnote{The dependence of the final abundance on the relativistic degrees of freedom $g_{eff}$ is mild unless freeze-out occurs near the QCD phase transition where $g_{eff}$ changes abruptly. Note as well that we use $g_{eff}(T)$ from~\cite{Laine:2006cp}.}.  More generally however, if DM is not self-conjugate, DM number violation may lead to a nonzero particle/antiparticle asymmetry, $\eta_{X} \equiv (n_{X}-n_{\overline{X}})/s$, where $s$ is the entropy density, $n_{X}$ is the particle number density and $n_{\overline{X}}$ is the antiparticle number density. In this case, the final abundance depends on the DM mass, annihilation cross section, and the asymmetry $\eta_{X}$.  

The Boltzmann equations for the evolution of $X$ and $\overline{X}$ in this more general setting are 
\be \frac{dn_{i}}{dt} + 3 H n_{i} = - \langle \sigma_{ann} v_{rel} \rangle \left[ n_{i}n_{j} - n_{eq}^2\right],
\ee
where the indices run over $i,j = X, \overline{X}$, $H$ is the Hubble expansion rate, and $n_{eq}$ is the equilibrium number density. Throughout we adopt the standard temperature parameterisation of the annihilation cross section, $ \langle \sigma_{ann} v_{rel} \rangle = \sigma_{0} \left( T/m_{X}\right)^{n}$, such that $n=0$ and 1 for $s-$ and $p-$wave annihilation respectively.

\begin{figure}[b]
\begin{center}
\includegraphics[width=.45\textwidth, height=0.45\textwidth]{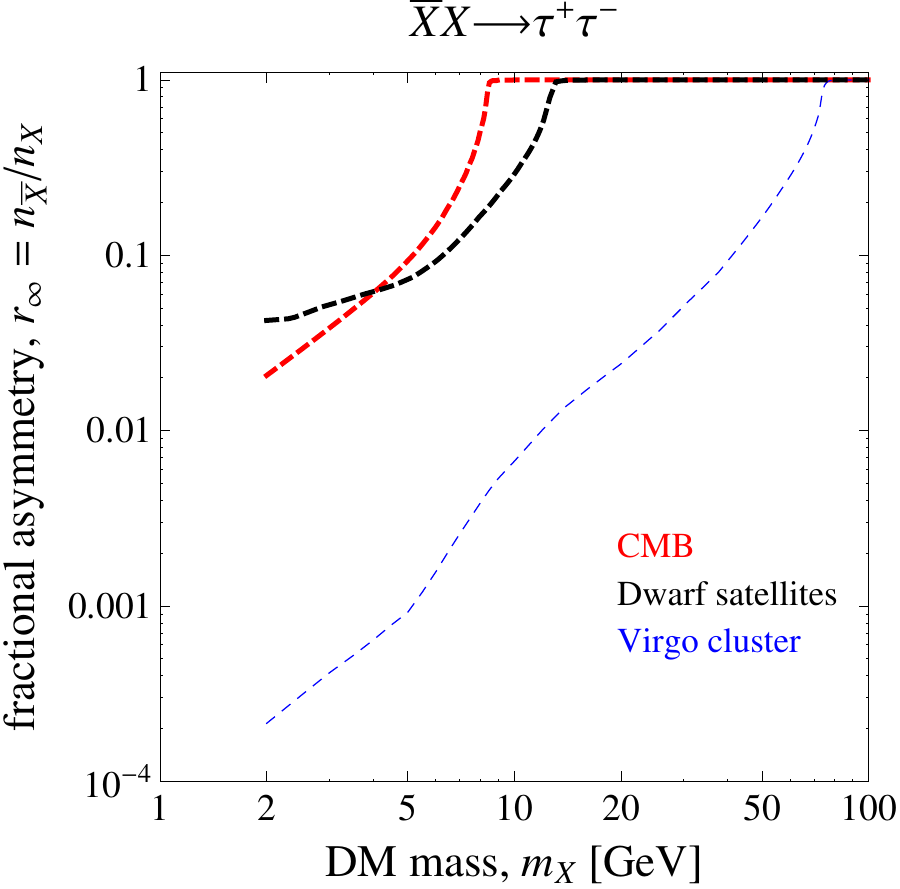}
\caption{CMB (bold red), dwarf satellite (bold black), and Virgo cluster (thin blue) limits on the fractional asymmetry $r_{\infty} \equiv n_{\overline{X}} /n_{X}$ for thermal ADM. For illustration we have fixed the annihilation channel to 100$\%$ $\tau^{+}\tau^{-}$. As in Fig.~\ref{fig:limits} bold lines denote robust limits, while the thin lines denote limits which are subject to larger systematic uncertainties.
}
\label{fig:rinf}
\end{center}
\end{figure}

\begin{figure*}[t]
\begin{center}
\includegraphics[width=.4\textwidth, height=0.4\textwidth]{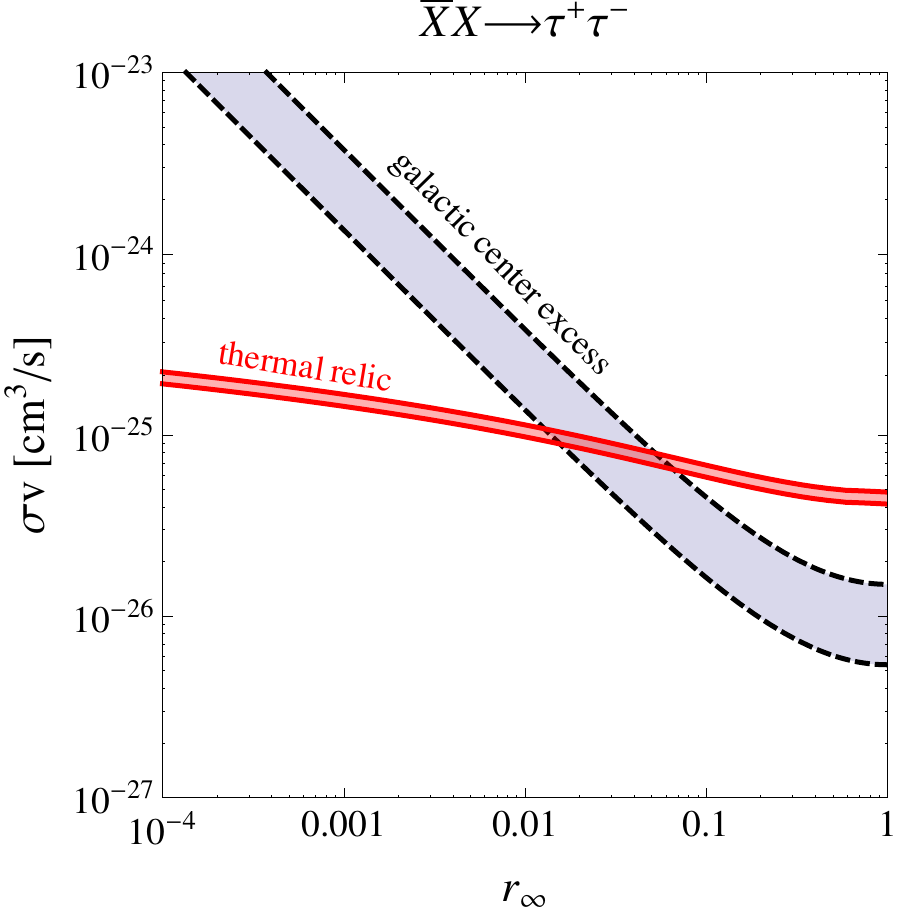}\includegraphics[width=.4\textwidth, height=0.4\textwidth]{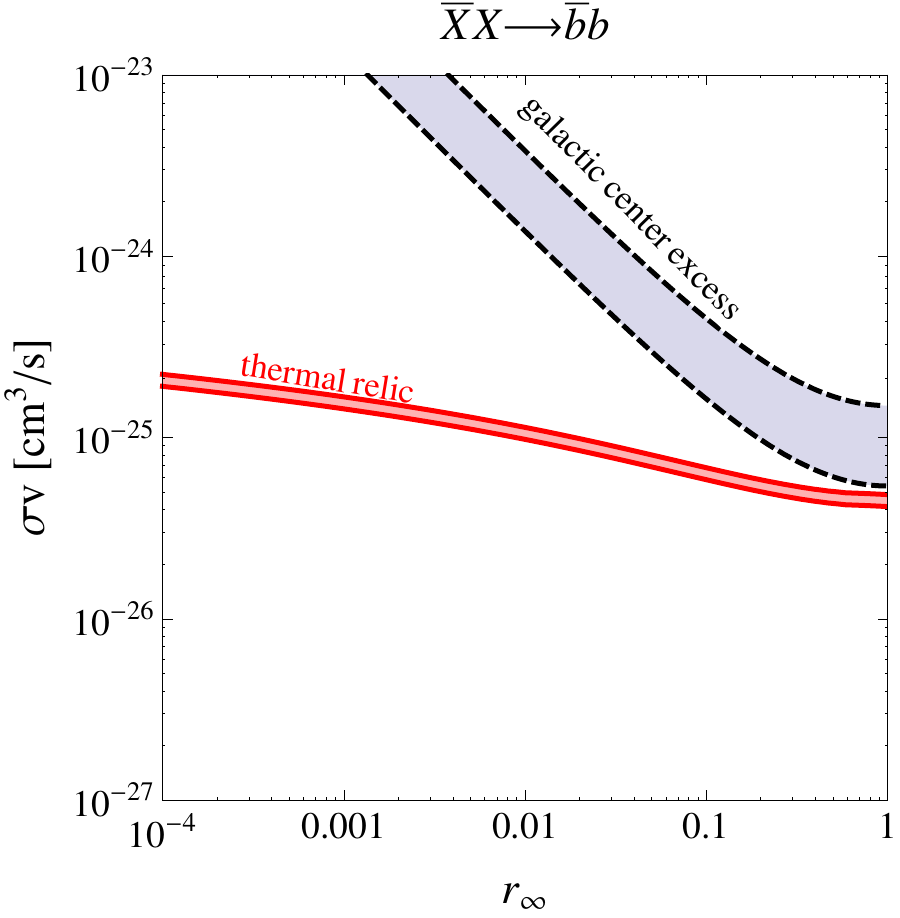}
\caption{The region of interest for the GC excess (shaded blue) for ADM annihilating into $\tau^{+}\tau^{-}$ (left) and $b\overline{b}$ (right) along with the thermal relic curve (red). Here we have varied the DM mass within the uncertainty, $39.4^{+3.7}_{-2.9}({\rm stat.}) \pm 7.9 ({\rm sys})$ GeV for the $\overline{b}b$ channel and $9.43^{+0.63}_{-0.52} ({\rm stat.}) \pm1.2 ({\rm sys.})$ GeV in the $\tau^{+}\tau^{-}$ channel~\cite{Abazajian:2014fta}.}
\label{fig:GC}
\end{center}
\end{figure*}

 A very accurate analytic solution to the above system of equations can be found in terms of the relic {\it fractional asymmetry}~\cite{Graesser:2011wi}
\be
r_{\infty} \equiv \frac{n_{\overline{X}}} {n_{X}} = \exp \left( \frac{ -\eta_{X} \lambda \sqrt{g_{*}}}{x_{f}^{n+1} \left(n+1\right)}\right),
\label{eq:rinf}
\ee
where $x_{f} \equiv m_{X}/T_{f}$ with $T_{f}$ the freeze-out temperature, $\lambda = 0.264 M_{Pl} m_{X} \sigma_{0}$, with $\langle \sigma_{ann} v_{rel} \rangle = \sigma_{0} (T/m_{X})^n$. The quantity $\sqrt{g_{*}}$ is a convenient combination of the entropic $h_{eff}$ and energy $g_{eff}$ degrees of freedom, $\sqrt{g_{*}} \equiv \frac{h_{eff}}{\sqrt{g_{eff}}} \left(1 + \frac{T}{3 h_{eff}} \frac{d h_{eff}}{dT} \right)$. The freeze-out temperature is found in the standard way as $x_{f} \simeq \log \left[ (n+1) \sqrt{g_{*}} a \lambda \right]$.

We can use Eq.~(\ref{eq:rinf}) to accurately compute the relic abundances of the $X$ particle and anti-particles, $\Omega_{X}$ and $\Omega_{\overline{X}}$, as 
\be  
\Omega_{DM} = \Omega_{X} + \Omega_{\overline{X}}= \frac{m_{X}s_{0} \eta_{X}}{\rho_{c}}\left(1 + \frac{2 r_{\infty}}{1-r_{\infty}}\right),
\\ \nonumber
\ee
where $s_{0}$ is the present-day entropy density.

When the fractional asymmetry is small $r_{\infty} \ll 1$, we can safely neglect the abundance of $\overline{X}$'s and write $\eta_{X} \simeq \rho_{c} \Omega_{DM} /(m_{X} s_{0})$, where we have introduced the critical density $\rho_{c}$ and the current entropy density $s_{0}$.  Combining this with Eq.~(\ref{eq:rinf}) gives an approximate value for the requisite asymmetric thermal annihilation cross section  asymmetric DM~\cite{Lin:2011gj} 
\be 
\langle \sigma_{ann} v \rangle_{ADM} \simeq \sqrt{\frac{45}{\pi}} \frac{(n+1)x_{f}^{n+1} s_{0}}{\rho_{c} \Omega_{DM} M_{Pl} \sqrt{g_{*}}} \log \left(\frac{1}{r_{\infty}}\right).
\label{eq:approx}
\ee
It can be seen that Eq.~(\ref{eq:approx}) agrees rather well with the numerical solutions shown in Fig.~\ref{fig:rinfrelic}.  We note that the results of Fig.~\ref{fig:rinfrelic} agree quite well with~\cite{Steigman:2012nb} with the proviso that these authors studied self-annihilating DM.

With these requirements for a consistent thermal relic scenario for ADM we now turn to the annihilation constraints for ADM.

\subsection{Constraints on the Fractional Asymmetry}

Let us now describe a simple method for re-expressing limits derived on symmetric annihilating DM as limits on asymmetric DM. Note that most annihilation limits are derived under the assumption that DM is self-annihilating, as appropriate for a self-conjugate field such as a Majorana fermion, whereas ADM models contain complex fields such as Dirac fermions or complex scalars.

With this difference in mind we can write the ratio of the ADM annihilation rate to the symmetric, self-annihilating case at a given DM mass as\footnote{Note that a similar expression in Ref.~\cite{Graesser:2011wi} has an overall factor of 2 difference from the present formula Eq~(\ref{eq:rate}), arising from the fact that there it was a ratio of Dirac ADM to Dirac SDM.} 
\be  
\frac{R_{ADM}}{R_{SDM}} = \frac{\langle \sigma v \rangle_{ADM}}{\langle \sigma v \rangle_{SDM}} \frac{r_{\infty}}{2}\left( \frac{2 }{1+ r_{\infty}}\right)^{2} 
\label{eq:rate}
\ee
where the ADM annihilation rate $R_{ADM} = n_{X} n_{\overline{X}} (\sigma v)$.  Eq~(\ref{eq:rate}) was derived by rewriting the ADM annihilation rate $R_{ADM} = n_{X} n_{\overline{X}} (\sigma v)$, using $r_{\infty} = n_{\overline{X}}/n_{X}$ and $2 n_{SDM} = n_{X} (1+ r_{\infty})$ where $n_{SDM}$ the number density of DM in the symmetric case.

As a case in point, let us illustrate the applicability of Eq.~(\ref{eq:rate}) by applying it to CMB limits (for previous WMAP7 limits, see~\cite{Lin:2011gj}). The current (WMAP9+Planck+ACT+SPT+BAO+HST+SN) 95$\%$ CL limit on symmetric DM energy can be simply recast as an ADM limit as
\bea 
f  \frac{ \langle \sigma v\rangle_{ADM}}{m_{X}} \frac{r_{\infty}}{2} && \left( \frac{2 }{1+ r_{\infty}}\right)^{2} \\ && < 1.2\times10^{-27}~{\rm cm}^{3}{\rm s}^{-1}{\rm GeV}^{-1}. \nonumber
\label{eq:CMB}
\eea
%
%
Making the additional assumption that the relic abundance is set by thermal freeze-out, we can combine Eq~(\ref{eq:CMB}) with Eq.~(\ref{eq:approx}) to determine a limit on the fractional asymmetry
\bea 
r_{\infty} && \log\left(\frac{1}{r_{\infty}}\right)  \nonumber \\
&& < f^{-1}~1.2\times10^{-2} \left(\frac{20}{x_{f}}\right) \left(\frac{m_{X}}{1{\rm GeV}}\right) \sqrt{\frac{g_{*}}{16}}. 
\eea
We plot the resulting constraint on $r_{\infty}$ in Fig.~\ref{fig:rinf}.

\subsection{Annihilation Signals in ADM}

Alternatively, when faced with potential signals of annihilating DM we can interpret the results as positive as information about the preferred values of $\langle \sigma v \rangle$ and $r_{\infty}$. As a case in point, we shall examine the extended gamma-ray emission from the Galactic Center (GC) of the Milky Way which has been interpreted as possibly arising from DM annihilation~\cite{Daylan:2014rsa,Abazajian:2014fta}.

To estimate the favored region in an ADM framework, let us fix the DM mass to the best-fit value for a given annihilation channel. Then with the use of Eq.~(\ref{eq:rate}) we plot in Fig.~(\ref{fig:GC}) the GC favored parameter space along with the thermal relic curve. Interestingly an ADM thermal relic can account for the GC excess with $r_{\infty} \sim 1$ for annihilation into $b$ quarks and $r_{\infty} \sim 0.02$ for annihilation into $\tau$ leptons.

\section{Discussion}

In the near-term future, substantial improvements in the sensitivity to DM annihilation are expected. In addition to increased data, the dwarf gamma-ray limits will benefit from future dwarf galaxy discoveries. Data from the Sloan Digital Sky Survey have roughly doubled the number of known MW satellites despite covering only $\sim 25$\% of the sky \cite{Willman:2009dv}. New generations of deep, wide-field photometric surveys (e.g., PanSTARRS, Dark Energy Survey, and Large Synoptic Survey Telescope) will potentially significantly increase the number of dwarf galaxies that can be used for DM searches, although estimates vary \citep{Tollerud:2008ze,Hargis:2014kaa}. 

\begin{figure}[t]
\begin{center}
\includegraphics[width=.45\textwidth, height=0.45\textwidth]{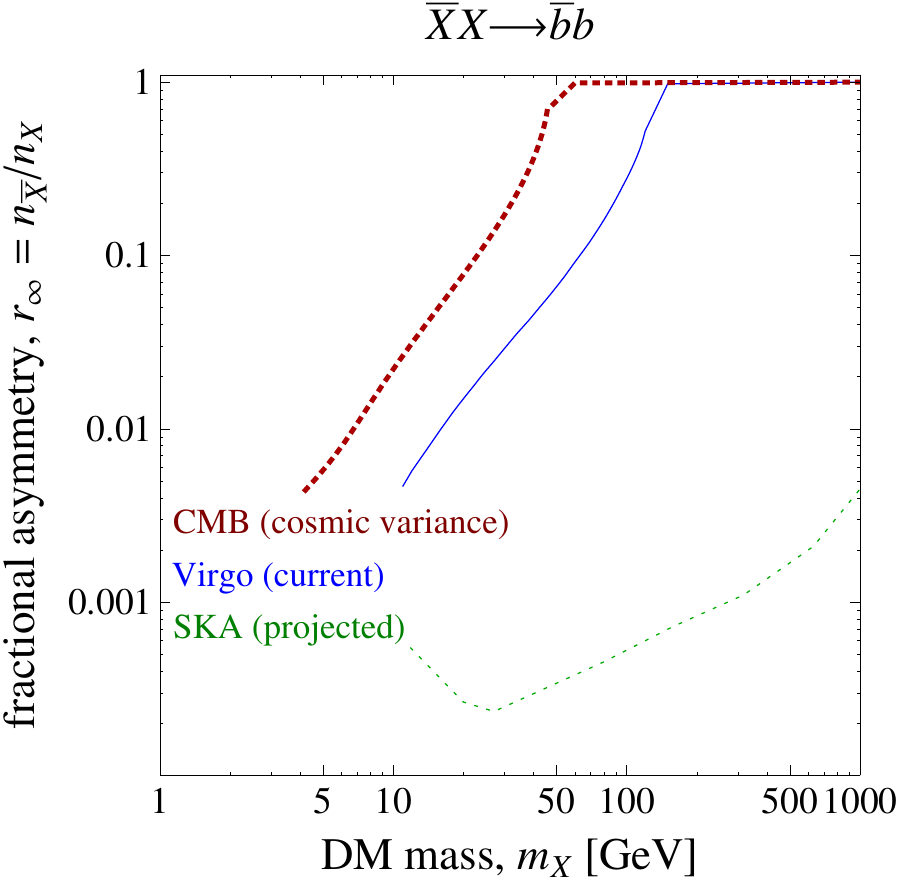}
\caption{CMB cosmic variance limited experiment (dark red, bold dotted), current Virgo cluster (blue, thin solid), and projected SKA radio (green, thin dotted) limits on the fractional asymmetry $r_{\infty} \equiv n_{\overline{X}} /n_{X}$ for thermal ADM. For illustration we have fixed the annihilation channel to 100$\%$ $\overline{b}b$. Bold lines denote robust limits, while the thin lines denote limits which are subject to larger systematic uncertainties.}
\label{fig:proj}
\end{center}
\end{figure}

It has also been argued that radio synchrotron emission from DM annihilation in dwarf satellites can provide strong limits~\cite{Regis:2014tga}. In particular, future detection prospects are especially promising for the Square Kilometre Array (SKA), which once completed will be the largest radio telescope on Earth. The resulting limits on DM annihilation are however subject to large uncertainties in the dwarf interstellar medium primarily related to their poorly known magnetic properties. In order to illustrate a rough future constraint from SKA we use the ``average'' benchmark model from~\cite{Regis:2014tga} (see their Table II) for annihilation into $\overline{b}b$. This projected limit has sub-thermal relic sensitivity for DM masses in the 10 GeV - 5 TeV range. Note that in more pessimistic models about dwarf gas content, magnetic fields, and diffusion properties, the projected cross section limits are roughly two orders of magnitude weaker~\cite{Regis:2014tga}.

These limits can be compared against the present Virgo cluster~\cite{Han:2012uw} and future CMB limits. For the latter we adopt for illustrative purposes the cosmic variance limited sensitivity, $\langle \sigma v \rangle /m_{X} \le 1.5 \times 10^{-28}~{\rm cm}^{3}{\rm s}^{-1}{\rm GeV}^{-1}$~\cite{Madhavacheril:2013cna}. We plot these limits together in Fig.~\ref{fig:proj}. There we observe that the CMB sensitivity can at best reach the $r_{\infty} \sim 4\times 10^{-3}$ level, while the current Virgo data already supersedes this. The projected SKA limits however cut substantially into ADM parameter space, requiring $r_{\infty} \lesssim 7\times10^{-4}$ for DM masses in the 10 GeV - 100 GeV range. We emphasize that future data limiting DM annihilation may not ever rule out a thermal relic, though it can be used quantify the degree of particle-antiparticle asymmetry required to render the thermal relic hypothesis compatible with observations.  

Lastly we observe that the preferred values of $r_{\infty}$ and $m_{X}$ from the GC and thermal relic requirements can be used to determine the ADM primordial asymmetry, $\eta_{X}$. The DM abundance in ADM can be written as $\frac{m_{X}\eta_{X}}{m_{p}\eta_{B}} = \left(\frac{1-r_{\infty}}{1+r_{\infty}}\right) \frac{\Omega_{DM}}{\Omega_{B}}$. Thus we observe that the $\tau$-channel requires $\eta_{X} \simeq (0.51-0.43) \eta_{B}$, while the $b$-channel is consistent with $\eta_{X} \lesssim 0.02 \eta_{B}$.

\section{Conclusions}

In this paper we have considered the model-independent consequences of indirect annihilation signals on thermal asymmetric DM.  We derived constraints on thermal ADM arising from the CMB, as well as dwarf satellite, and Virgo cluster gamma-ray data from the Fermi-LAT satellite. In contrast with what one may naively assume for ADM, annihilation constraints can be strong. 
%
%
These are especially stringent for light DM where, for example, the
CMB and dwarf satellite data require $n_{\bar{X}}/n_{X} < 0.07$ for
$m_{X} < 5$ GeV with 100$\%$ annihilation into $\tau^{+}\tau^{-}$. The
Virgo results imply the even more stringent requirement
$n_{\bar{X}}/n_{X} < 0.001$ for $m_{X} < 5$, but are subject to
greater uncertainties.

In addition we have determined the preferred fractional asymmetry $r_{\infty} \equiv n_{\bar{X}}/n_{X}$ from the extended source of gamma-rays from the GC. The $\bar{b}b$ annihilation channel is in mild tension with the requirements of a thermal relic, but at $> 2 \sigma$ is consistent with $r_{\infty} \sim 0.5 -1$. In contrast, the $\tau^{+}\tau^{-}$ annihilation channel is rendered fully consistent with thermal relic considerations for $r_{\infty} \sim 0.01 - 0.04$. Thus in contrast with the self-annihilating case, thermal ADM is compatible with the Galactic Center annihilation rates for both $\overline{b}b$ and $\tau^{+}\tau^{-}$ annihilation channels. We note lastly that though all DM interpretations of the Galactic Center excess have tension with Virgo cluster limits in~\cite{Han:2012uw}, this could be simply resolved if boost factors are closer to those estimated in~\cite{Sanchez-Conde:2013yxa}. 

\vspace{.5cm}

\acknowledgements

We are very grateful to Kevork Abazajian, John Beacom, and Manoj Kaplinghat for useful comments.  We would also like to thank Mikko Laine for sharing the data on $g_{eff}(T)$ from~\cite{Laine:2006cp}, Marco Regis for sharing the SKA projection limits from~\cite{Regis:2014tga}, and Tuomas Karavirta for very helpful computing advice.  The CP$^3$-Origins center is partially funded by the Danish National Research Foundation, grant number DNRF90.

\section*{Appendix}

For reference we have used $s_{0} = 2878~{\rm cm}^{-3}$, and $\Omega_{CDM} \rho_{c} = \left(\Omega_{CDM} h^{2}\right)~1.054\times10^{4}~{\rm eV}~{\rm cm}^{-3}$, with the best-fit Planck result $\Omega_{CDM}h^{2} = 0.12029$~\cite{Ade:2013zuv}.

\bibliography{asym_references.bib}

\end{document}